\documentclass{article}%
\usepackage{amsmath}
\usepackage{amsfonts}
\usepackage{amssymb}
\usepackage{graphicx}
\usepackage{multirow}
\usepackage{fleqn}

\title{Numerical Studies of Hamiltonian Systems and Application to Galactic Potentials}
\author{Daniel Pfenniger\\
Geneva Obstervatory, Switzerland}
\begin{document}

\maketitle{}

\section{Talk Summary}\label{sec:intro}
The talk consisted mainly in commenting in a linear way the H\'{e}non \& Heiles (1964) paper.  Instead of repeating here the lecture of
the paper, we advise the reader interested in dynamical systems to study this ``must'' reading.  Below are a few comments added
during the talk and references.
   
Michel H\'{e}non's contact with Geneva Observatory started in the 70's when the third course of the Saas Fee series was organized by
Louis Martinet and Michel Mayor including Michel H\'{e}non as speaker (H\'{e}non 1973), together with Donald Lynden-Bell and Georges
Contopoulos.  Around this time Louis Martinet, my thesis supervisor, inspired by Michel H\'{e}non worked at understanding the chaos in
galactic potentials (e.g., Martinet 1974), which led naturally to my thesis topic, the dynamics of barred galaxies, which are
typical systems where chaos and regular motion coexist each in substantial parts.
 
All this activity was for good part consequence of the seminal paper in 1964 by Michel H\'{e}non and graduate student Carl Heiles at
Princeton University: \textit{``The applicability of the third integral of motion: Some numerical experiments''} in the field of
galactic dynamics.  Although the incentive is galactic dynamics, the scope of this paper is much broader.  Together with the paper
of Edward Lorentz ``Deterministic Nonperiodic Flow'' (1963) these two papers were pivotal in launching a new field of research
called ``dynamical system theory''; H\'{e}non \& Heiles paper for the domain of Hamiltonian dynamical systems, and Lorentz' paper for
the domain of dissipative systems.  Both papers were motivated by modeling concrete scientific questions with computers (the third
integral in galactic potentials and the long-term weather prediction, respectively). Before these papers, dynamical systems were
widely believed to be either completely integrable or completely ergodic.  After, the possibility of \textit{semi-ergodic} motion (a
very appropriate adjective used by H\'{e}non but later replaced in popularity by ``chaotic'') provided a continuum of possibilities
between the two integrable or ergodic extreme cases, which represent actually a measure zero subset in the set of all dynamical
systems.  Chaos is characterized by sensitive dependence on initial conditions and also often occurs close to resonances.  As argued
in H\'{e}non \& Heiles paper, if in a given dynamical system a range of initial conditions show a transition from regular quasi-periodic
to chaotic motion occurs, often the transition is sharp.

The H\'{e}non \& Heiles paper was promoting the use of computers to perform numerical experiments as a proper research method. An
earlier example in this area of this kind was the famous Fermi, Pasta \& Ulam paper (1955) trying precisely to understand the
transition from quasi-periodic to ergodic in a 1-dimensional chain of non-linear oscillators.  The H\'{e}non \& Heiles paper was also
promoting the use of the clever tool of the surface of section in dynamical systems, invented by Poincar\'{e} long before (1899) but not
much used in numerical works.  H\'{e}non brought a step further the application of this tool, replacing the time-continuous Hamiltonian
system by a time-discrete iterated Hamiltonian mapping, gaining by a factor $\sim 1000$ in computing power if the objective is just
to study \textit{typical}, in this case, Hamiltonian systems.  I used precisely this trick in my thesis for understanding the phase
space neighborhood of complex unstable periodic orbits, first by using 4D Hamiltonian mappings and then applying the knowledge to
galactic orbits (Pfenniger 1985ab).

At the same time, the mathematical understanding of Hamiltonian systems increased much through the KAM theorem (Kolmogorov 1954;
Arnold 1963; Moser 1962).  By using numerical experiments Michel H\'{e}non brought much clarity fo the KAM theorem meaning for concrete
dynamical problems to the physicical and astronomical communities.  Without such an insight brought by computer plots, probably the
KAM results would have stood confined to the mathematical community for many decades, very much like the multi-decade old idea of
fractal sets which was popularized some years later by Benoît Mandelbrot (e.g., Mandelbrot 1977) using also simple computer models
and graphical representations.

The H\'{e}non \& Heiles paper, currently with about 1000 citations, is the most cited paper of Michel H\'{e}non's list, due to its breath and
innovative content touching fundamental aspects of classical mechanics. It reached its citation rate peak in 1985, and since then
the citation rate slowly decreases, which characterizes a pioneer and seminal paper, two decades in advance over the scientific
community main preoccupations, and  still of lasting value 50 years later.

My only curiosity frustration bears on the background work methods unexplained in the paper, probably for the sake of clarity.  For
example, several numerical or procedural clever methods had to be invented with the early 60's computers, in particular to elaborate
the plots.  For instance, the elegant method for cleanly computing surface of sections was only explained much later by H\'{e}non (H\'{e}non
1982), although computer codes including it were circulating well before.
 
In conclusion, the H\'{e}non \& Heiles paper is exemplary in many ways and introduce several innovations.  It is clear, concise and
pedagogical.  It is exemplary especially on the method used by H\'{e}non throughout his research: to simplify a problem as much as
possible to gain in generality, yet keeping the non-trivial properties that make the problem difficult.

\section*{Bibliography}

\begin{itemize}

\item Arnold, V.I. 1963a, \textit{Proof of a Theorem by A. N. Kolmogorov on the invariance of quasi-periodic motions under small
    perturbations of the Hamiltonian}. Russian Math. Survey 18 : 13-40.

\item Fermi, E.; Pasta, J.; Ulam S. 1955, \textit{Studied of non linear problems}, Los Alamos Document LA-1940

\item H\'{e}non, M. 1973, in \textit{Dynamical Structure and Evolution of Stellar Systems}, Saas-Fee Advanced Course 3, Swiss Society
  for Astronomy and Astrophysics (SSAA). Edited by L. Martinet \& M. Mayor. Observatoire de Genève, Switzerland, p. 183
	
\item H\'{e}non, M. 1982, \textit{On the numerical computation of Poincar\'{e} maps}, Physica D: Nonlinear Phenomena, 5, 412-414.
	
\item H\'{e}non, M.; Heiles, C. 1964, \textit{The applicability of the third integral of motion: Some numerical experiments},
  Astronomical Journal, 69, 7

\item Kolmogorov, A.N. 1954, \textit{On the conservation of conditionally periodic motions under small perturbation of the
    Hamiltonian}. Dokl. Akad. Nauk. SSR 98: 527-530.

\item Mandelbrot,  B. 1977, \textit{Fractals: Form, Chance and Dimension}, W H Freeman and Co

\item Martinet, L. 1974, \textit{Heterclinic Stellar Orbits and ``Wild'' Behaviour in our Galaxy}, Astronomy and Astrophysics, 32,
  329

\item Moser, J.K. 1962, \textit{On invariant curves of area-preserving mappings of an annulus}. Nach. Akad. Wiss. Göttingen,
  Math. Phys. Kl. II 1 : 1-20.
	
\item Pfenniger, D. 1985a, \textit{Numerical study of complex instability. I - Mappings}, Astronomy and Astrophysics 150, 97

\item Pfenniger, D. 1985b, \textit{Numerical study of complex instability. II - Barred Galaxy Bulges}, Astronomy and Astrophysics 150, 112

\item Poincar\'{e}, H. 1892, \textit{Les M\'{e}thodes Nouvelles de la M\'{e}canique C\'{e}leste}. Paris: Gauthier-Villars. 

\end{itemize}

\end{document}